\documentclass{iopart}
\usepackage{graphicx}
\usepackage{bm}
\begin{document}
\review{Coarse grained models in Coulomb-frustrated phase separation}
\author{C. Ortix $^1$, J. Lorenzana $^{2,3}$ and  C. Di Castro $^3$}
\address{$^1$ Dipartimento di Fisica, Universit\`{a} del Salento and INFN Sezione di Lecce, Via per Arnesano, 73100 Lecce, Italy.}
\address{$^2$ ISC-CNR, Via dei Taurini 19, 00185 Roma .}
\address{$^3$ SMC, INFM-CNR, Dipartimento di Fisica, Universit\`a di Roma 
``La Sapienza'', P.  Aldo Moro 2, 00185 Roma, Italy.} 
\ead{carmine.ortix@le.infn.it}
\begin{abstract}
Competition between interactions on different length scales leads to
self-organized textures in classical as well as quantum systems. This
pattern formation phenomenon has been invoked to explain some
intriguing properties of a large variety of strongly correlated
electronic systems that includes for example high temperature
superconductors and colossal magnetoresistance manganites. We classify
the more common situations in which Coulomb frustrated phase
separation can occur and review their properties. 
\end{abstract}

\submitto{\JPCM}

\section{Introduction}
\label{sec:intro}

A large variety of systems with competing short and long range
interactions self organizes in domain patterns\cite{seu95}.  
Examples  range from  ferromagnetic systems\cite{kit46,lan84}
 to diblock copolymers\cite{oht86}  and, 
at least theoretically,  neutron star matter\cite{cplor93}. 
Inhomogeneous states display  a simple set of
predominant morphologies like circular droplets
and stripes in two-dimensional (2D) systems, and layers,  cylindrical rods and
spherical droplets in three-dimensional (3D)
systems. This tendency for a common behavior across different systems
calls for simple models  which neglect the specific details of each
system and capture the general properties.

In this work we will discuss phase separation frustrated by the 3D
Coulomb interaction in electronic systems but similar ideas apply
also to neutron star matter and diblock copolymers. 
Tendency to phase separation in electronic systems manifests itself by
anomalies in the electronic contribution to the free energy density $f_{e}$. 
Two kind of anomalies appear often in strongly correlated systems. 
The  first situation is determined by a
negative compressibility region. A notorious example is the uniform
electron gas (EG) at low density\cite{mah00} but this feature appears in
several other 
models\cite{nag67,mar90,cas95b,don95,kag99,lor01II,fre02,kug05,fis07}
including neutron star matter\cite{cplor93}.

The other possibility is that the inverse electronic
compressibility has a point with a Dirac-delta-like
negative divergence at some density $n_c$.
This happens when the free energies of two states
which are separated by a barrier, cross each other leading to a cusp
singularity. An example is also provided  by the EG. Indeed numerical
simulations show that the Wigner crystal and the uniform phases free
energy cross at some density\cite{cep80}. The same feature
appears in several models including
manganites\cite{bri99b,oka00}.

The behavior of
$f_e$   can be summarized by
expanding the electronic free energy density $f_e\equiv F_e/V$ around a reference density $n_{c}$
as $f_e = \alpha |n-n_{c}|^\gamma$ with $\alpha<0$ and correspond to
$\gamma=2$ (negative compressibility region) or  $\gamma=1$ (cusp
behavior). A neutral system with $\alpha<0$ is unstable toward phase
separation however in a charged system, this tendency is frustrated by
the long range Coulomb interaction. This leads to the formation of
inhomogeneities with a characteristic size $l_d$  determined by the 
competition between long range forces and surface energy effects.

Regarding the experiments, the first task is to determine what can be 
attributed to Coulomb frustrated phase separation and what can not. 
Information on the mechanism of the segregation can be obtained by an
observation of the morphology. 
For example in magnetoresistant manganites, inhomogeneous states with
large insulating and/or metallic clusters 
with very rough interfaces have been reported\cite{fat99}. 
 These resemble the domains of the random field Ising model where the surface
energy plays a role only at very short length scales. The shape of the
interfaces is determined by the fluctuations of the local field which
stabilizes one or the other of two competing phases and long range
forces are not needed\cite{dag01,bur01}. 

A quite different morphology has been observed, also in manganites, by
scanning tunneling spectroscopy in thin films.  Domains have a
filamentary and droplike metallic and/or insulating regions
in the scale of tens to thousands of nanometers with smooth surfaces
indicating strong surface energies \cite{bec02} which is consistent
with Coulomb frustrated phase separation as predicted by Nagaev
several years ago\cite{nag83}. This is reinforced by the fact that the 
morphology is similar to that of classical mesodomains\cite{seu95}.   
Although  quenched disorder can have an
important role altering the ideal periodic domain configurations 
that theories would predict in the clean limit, the strong
dissimilarity with the random field Ising model domain morphology 
makes it clear that quenched disorder is
not the driving force. In this regard it is interesting that, even in
the absence of quenched disorder, the complexity of the energy
landscape of frustrated phase separation models can make the  
ordered ground state unreachable leading to 
a glassy state\cite{sch00}.

It is not clear at the moment which of the
two mechanism (quenched disorder or short and long range forces)
is dominant to determine each of the anomalous
properties of complex electronic systems and more theoretical and
experimental work is needed. Several recent experiments, however, point 
to an important role of the  short and long range force competition. 
In the high-temperature superconductivity context, scanning tunneling microscopy investigations \cite{lan02} have revealed an inhomogeneous state of electronic  superconductinglike nanopatches  coexisting with poorly metallic regions. The
 domains have roundlike shapes indicating that the surface energy has
 an important role which points to a Coulomb frustrated phase
 separation mechanism.

Recently a strong transport anisotropy in ultra clean
Sr$_{3}$ Ru $_{2}$O$_{7}$ samples have been observed \cite{bor07}. This
striking feature is consistent with the proposal of exotic electronic
liquid phase analogue \cite{kiv98} to the intermediate order states of
liquid crystals \cite{cha95}. 
The idea that the domains in Sr$_{3}$ Ru $_{2}$O$_{7}$ are due to 
frustrated phase separation has already been put
forward\cite{hon05}.

The phenomenology of ruthanates has a strong similarity to
that observed in GaAs heterostructures. Several magneto-transport
anisotropies arise when the Fermi level lies near the middle of a
highly excited Landau level \cite{lil99}. It has been proposed
\cite{kou96} that in a clean two-dimensional electron gas (2DEG) in
high Landau levels, a uniform phase would be unstable against a charge
density striped phase where the electron density alternates between
zero and full-filling.

Evidences for inhomogeneous states have also been reported in the 2DEG
at zero magnetic field. Using a local probe, Ilani and collaborators
\cite{ila01,ila00} have shown that close to the still debated
Metal-Insulator transition (MIT), mesoscopic inhomogeneities
appear. In addition thermodynamic measuraments have shown that close
to the MIT the compressibility departs sharply from the predictions of
an homogeneous electron gas\cite{eis92,eis94}. Another interesting
finding is the appearance of negative spikes in the electronic
compressibility \cite{ila01} which indicate that the transition from
the homogeneous state to the inhomogeneous state is discontinuous as
found theoretically for a striped Coulomb frustrated phase separated
state\cite{ort06}.

In cuprates, as in manganites, the situation is more complex and
different inhomogeneities have been reported at different length
scales. At a scale of  $\sim 10nm \sim 20$ lattice
constants  the system appears to 
segregate into a pseudogap or underdoped  phase with a large gap and a 
superconducting phase with a smaller gap with smooth interfaces 
in between\cite{lan02}. This is consistent with Coulomb frustrated
phase separation between an underdoped pseudogap phase and an 
overdoped phase\cite{lor02b}. This scenario has been reproposed  on the
light of recent neutron scattering experiments\cite{wak07}.
At a scale of $\sim 4$ lattice constants the systems
shows charge ordering in the form of stripes\cite{tra95}. 
By considering the CuO$_2$ plane as a strongly correlated 2D
electronic gas coupled to the lattice, in the presence of the
long-range Coulomb interaction, it has been proposed\cite{cas95b}
that an incommensurate charge density wave occurs around optimal
doping as Coulomb frustrated phase separation with hole rich and hole
poor regions smoothly evolving intro stripes by underdoping.

Since domains in these system have often mesoscopic scales of several lattice 
constants, general aspect of the phenomena can be analyzed by the use of
coarse grained models. In this work we will review the two more common
coarse grained model corresponding to two different universality classes
and their differences and similarities. 
The paper is organized as follows: In Sec.~\ref{sec:models} we review
the more relevant models that along the years have been proposed for
$\gamma=1,2$ and introduce the basic length scales of the problem. 
In Sec.~\ref{sec:weaklambda} we discuss the main properties of
phase-separated states for weak Coulomb interaction where a unified
picture can be achieved. In Sec.~\ref{sec:stronglambda} we
discuss the strong interaction regime where strong differences between
$\gamma=1,2$ arise and we conclude in Sec.~\ref{conc}.

\section{Minimal models  and  typical length scales}
\label{sec:models}
As discussed in Sec.~\ref{sec:intro}, tendency to phase separation can
be sorted in two main electronic free energy anomalies, corresponding to a
negative compressibility density range or a Dirac-delta-like
singularity in the compressibility. Although both situations can be
captured expanding the free energy as $f_{e}\propto \alpha
|n-n_{c}|^{\gamma}$ with $\alpha<0$ and $\gamma=1,2$, a full 
analysis of the phase separation problem, from small to high
frustration, requires higher order terms in the free energy
expansion as follows:
\begin{center}
\begin{equation}
  \label{eq:feuniv}
f_{e}=\alpha|n-n_{c}|^{\gamma}+\beta |n-n_{c}|^{2 \gamma} . 
\end{equation}
\end{center}
 For $\beta>0$ this provides a double-well form for the free energy with minima at $n=n_{c} \pm
 \delta n^{0}/2$ (here $\delta n^0 \equiv 2 [|\alpha|/ (2
 \beta)]^{1/\gamma}$) and a barrier between minima of height $f_0=\alpha^2/(4\beta)$. 
This form of the 
 free energy assumes two symmetric phases with the same
 compressibility. One can consider asymmetric situations by adding 
cubic terms. Such asymmetry does not change the physics
 substantially and will be neglected for simplicity. 
The exception is the important case of an incompressible phase
coexisting with a compressible phase.  
This limiting case constitutes a
different universality class  and will not be discussed here. 
A treatment can be found in 
Ref.~\cite{lor02} in 3D and in Ref.~\cite{ort06} in 2D.

We start by introducing the $\gamma=2$ model which is defined by the
following free energy: 
\begin{eqnarray}
&F&=\int d^{D}x \left[\alpha \Delta n^{2}+\beta
        \Delta n^{4} +  c |\nabla  n\left({\bf x}\right)|^{2}\right] \\ 
&+&  \frac{Q^{2}}{2}\int d^{D}x \int d^{D}x^{\prime} 
\left[n\left({\bf x}\right)-\overline{n}\right]v({\bf x-x^{\prime}})\left[n\left({\bf x^{\prime}}\right)-\overline{n}\right]\nonumber 
\label{eq:eGL}
\end{eqnarray}
where $\overline{n}$ is the average charge representing the density of
the rigid background, $\Delta n=n-n_{c}$ and  $v=|{\bf
  x-x^{\prime}}|^{-1}$ is the Coulomb interaction. 
Finally  the gradient
term models the surface energy
of smooth interfaces and is parameterized by the stiffness constant $c$.
Here the charge density plays the 
role of a scalar order parameter in an  analogous way as the
liquid-gas transition of classical fluids. 
Pattern formation within this  model  has been
considered by Schmalian and
Wolynes\cite{sch00} in three-dimensional systems, and more recently by
Muratov \cite{mur02} who considered the case of D-dimensional systems
subject to a D-dimensional Coulomb interaction in the sharp-interface
limit. This model is also relevant to 
describe segregation in block copolymers\cite{oht86}.

Since Eq.~\eref{eq:eGL} has several  parameters, to study
the phase diagram it is convenient to  measure lengths 
in units of  $\xi=\sqrt{2 c/ \alpha}$, and define a dimensionless density  
 $\phi(x)= 2 \Delta n(x)
/ \delta n^{0}$ and  a dimensionless free
energy $\Phi\equiv F/(f_0 \xi^D)$  that, apart from an irrelevant constant reads:
\begin{eqnarray}
&\Phi & = \int d^{D} x \left[\phi({\bf x})^{2}-1\right]^2+|\nabla \phi({\bf x})|^{2}\label{eq:eGLrescale} \\
& +  & \frac{Q^2_{R}}{2 } 
\int d^{D}x\int d^{D}x^{\prime} \left[\phi \left({\bf x}\right)-
  \overline{\phi}\right]v\left({\bf x-x^{\prime}}\right)
\left[\phi \left({\bf x^{\prime}}\right)- \overline{\phi}\right]  \nonumber
\end{eqnarray}
with $ \overline{\phi}= 2(\overline{n}-n_{c})/ \delta n^{0}$ and $Q^2_{R}$ a rescaled dimensionless Coulomb coupling given by:
$$ Q^{2}_{R}=  Q^{2} \frac{2 \, \xi^{D-1}}{|\alpha|}.$$
We see that the parameter space can be reduced to only two parameters, the 
dimensionless global density $\overline\phi$ and the renormalized 
coupling $Q^{2}_{R}$. In three-dimensional systems $Q^{2}_{R}$
coincides with the dimensionless coupling introduced in
Ref.~\cite{mur02}.

The $\gamma=1$ case has been investigated  with approximate
treatments in three- \cite{lor01I,lor02} and two-dimensional systems
\cite{ort06} and within an exactly solvable model in the limit $\beta=0$
\cite{jam05} and $\beta\neq 0$\cite{ort07}.  
In this case frustrated phase separation 
is more easily described by adding
an auxiliary field $s$ linearly coupled to the charge and analogous to
a Hubbard-Stratonovich variable.  
Two versions of the model are possible which
lead to  similar results: $s$ can be taken as a soft or a conventional
Ising spin with $s=\pm 1$ where the sign distinguishes the two competing
phases. We take the latter model which is more
straightforward to analyze. It consists of a ferromagnetic Ising model
linearly coupled to the local charge: 
\begin{eqnarray}
F&=&-J \sum_{<i j >}(s_{i}s_{j}-1)-|\alpha|
\sum_{i} s_{i}\left(N_{i}-N_{c}\right) 
\label{eq:model} \\
& &  +\frac\beta{a^D} \sum_{i}\left(N_{i}-N_{c}\right)^{2}
 +\frac{Q^2}{2}\sum_{i j}\left(N_{i}-\overline{N}\right) v({\bf
 x}_i-{\bf x}_j) \left(N_{j}-\overline{N}\right)\nonumber
\end{eqnarray} 
where $s_i=\pm 1$, the index $i$ runs over the sites of a
hypercubic lattice of dimension $D=2,3$ with lattice constant $a$, the $N$'s are dimensionless numbers of particles per site and $\overline{N}$ their average value.
The soft version replaces the Ising part with  
a double well potential\cite{jam05}. 

\begin{figure}[tbp]
\begin{center}
\includegraphics[width=7 cm]{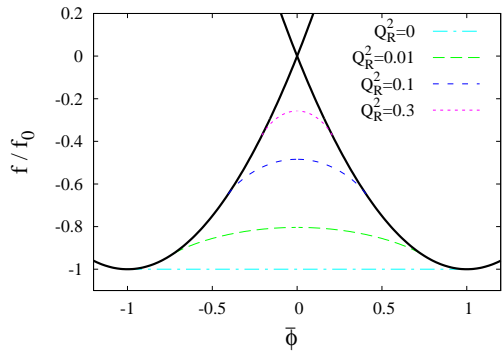}
\end{center}
\caption{Behavior of the free energy densities for the
  electronic uniform phases for the $\gamma=1$ model
  Eq.~\eref{eq:modelrescale} and $D=2$. The dashed lines are the exact
  energy of stripe inhomogeneities for
  different values of the frustrating parameter. The
  short-long dashed line 
  corresponds to the Maxwell construction ($Q^{2}_{R}=0$).}  
\label{energylambda}
\end{figure}

We have written the model in the
lattice for clarity but we are interested on the continuum limit of
this model with $n({\bf x})\equiv N_i/a^D$. Uniform phases correspond to a
ferromagnetic state in $s$. Inserting the two possible values of 
$s$ in Eq.~\eref{eq:model} one obtains that uniform phases are
described by Eq.~\eref{eq:feuniv} with $\gamma=1$, i.e. two
intersecting parabolas with minima at $\pm\delta n_0/2$ and a crossing
point at $n_c$ ( the full lines in Fig.~\ref{energylambda}). 
  In the hard spin case domain
walls of the Ising order parameter are sharp by construction with a
surface tension $\sigma= 2 J/a^{D-1}$ thus the Ising term can be
written as $\sigma \Sigma$ with $\Sigma$ the total surface of
interface among the two phases. It is convenient to define the
analogous of $\xi$ for the present model 
$\xi\equiv 4\sigma \beta/\alpha^2$. This represents the size that
inhomogeneities should have for the total interface energy 
be of the same order as the phase separation energy density gain 
$\alpha^2 /(4 \beta)$. As before we measure the energy in units of $
\xi^{D} \alpha^2 /(4 \beta)$, lengths in units of 
$\xi$ and surfaces in units of $\xi^{D-1}$. In these units
$\sigma\equiv1$ and apart from an irrelevant constant one obtains the
following free energy functional: 
\begin{eqnarray}
\Phi&=& \Sigma+\int d^{D} x [\phi({\bf x})-s({\bf x})]^{2} \label{eq:modelrescale}  \\
 & +  & \frac{Q_R^2}2 \int d^{D}x\int d^{D}x^{\prime}
 \left[\phi\left({\bf x}\right)- \overline{\phi}\right]v\left({\bf
 x-x^{\prime}}\right) 
\left[\phi \left({\bf x^{\prime}}\right)- \overline{\phi}\right]  
\nonumber
\end{eqnarray}
where 
$$Q_R^2=Q^2 \frac{\xi^{D-1}} \beta $$
As for the model Eq.~\eref{eq:eGLrescale} the parameter space is
determined by the two dimensionless parameters  $\overline{\phi}$ and
$Q^{2}_{R}$. Frustration can be also measured by the parameter
$Q^{2/D}_{R}$ which, as shown below, has the meaning of ratio between
the energy cost introduced by frustration and phase separation energy gain. This was the
parameter used in  Refs.~\cite{lor01I,lor02,ort06,ort07}.

In the absence of the Coulomb interaction, both models are subject to
ordinary phase separation in a range of global 
densities $|\overline{\phi}|<1$ as determined by the Maxwell (or
``tangent'') construction shown by the  short-long dashed line in Fig.~\ref{energylambda}. The
phase-separated state is made up of macroscopic domains with  constant local
densities $\phi=\pm 1$. 
For $Q^{2}_{R} \neq 0$, macroscopic phase separation is precluded
since the Coulomb cost grows faster than the volume in the
thermodynamic limit and mesoscale domains appear. 

The typical size of the domains $l_d$ can be obtained by dimensional
analysis. We define $\tilde l_d\equiv l_d/\xi$.
Taking $\phi\sim 1$ the  Coulomb cost per domain can be estimated
making the integrals in a volume of order $\tilde l_d^D$ as $Q_R^2 \tilde l_d^{2D-1}$. The surface energy
goes as $\tilde l_d^{D-1}$. Both quantities are optimized when the
inhomogeneities have the size  defined by
\begin{equation}
  \label{eq:ld}
\tilde l_d^D=1/Q_R^2.
\end{equation}

Another important length scale is the screening length of the Coulomb
interaction that can be defined for two- and three-dimensional systems
as: 
\begin{equation}
  l_{s}^{D-1}=\left[2^{D-1} \pi Q^{2} \kappa \right]^{-1}
\end{equation}
where $\kappa$ is a characteristic electronic compressibility
of the competing homogeneous  phases:
\begin{eqnarray}
  \label{eq:kappa}
  \kappa&=&(2\beta)^{-1} \;\;\;\;\;\ (\gamma=1)\\
  \kappa&=&(2|\alpha|)^{-1} \;\;\;\;\;\ (\gamma=2)
\end{eqnarray}
For both the models presented above the characteristic  screening
length in units of $\xi$ can be defined as
\begin{equation}
  \label{eq:ls}
\tilde l_{s}^{D-1} \equiv 1/Q^{2}_{R}.
 \end{equation}
At weak frustration $Q_R<<1$ we have the following hierarchy 
of scales\cite{mur02} [c.f. Eq.~(\ref{eq:ld}),(\ref{eq:ls})]:
\begin{equation}
  \label{eq:hierarchy}
l_s>>l_d>>\xi.
\end{equation}
This separation of lengths allows for a unified treatment of the
frustrated phase 
separation mechanism at weak frustration that will be discussed in the
following section.

\section{The weak frustration regime}
\label{sec:weaklambda}


 In the weak frustration regime, the effect of long-range forces can be
 considered as a small perturbation upon the ordinary phase separation
 mechanism.  
Thus, mixed states are expected 
to appear with local densities close to the two minima of the
double-well. 
For systems with $\gamma=2$, frustrated phase separation can be
analyzed by expanding quadratically the free energy around the two
densities $\phi=\pm 1$. Then the bulk free energy becomes the same as
for the $\gamma=1$ model. In addition, the hierarchy of
length scales Eq.~(\ref{eq:hierarchy}) for $Q^{2}_{R}<<1$ 
allows to consider the interface as sharp. A surface tension can be
defined by computing the excess energy of an isolated
interface\cite{mur02}. At this point, the two models become
equivalent. In the rest we present the analysis of the $\gamma=1$
model which can be solved analytically for stripes in $D=2$ and layers
in $D=3$. 

Before discussing the exact solution, it is convenient to discuss an
approximate solution which gives essentially the same result as the
exact solution in weak coupling but in addition allows to consider other
geometries. Since the state is expected to be similar to the
macroscopically phase separated state one can take a uniform density
approximation (UDA)
in which the local density inside the domains
is assumed constant\cite{lor01I,lor01II,lor02,ort06,ort07}. 
Comparison with the exact result shows 
that this gives very accurate results  in two- and three-dimensional systems 
\cite{ort07}. 

\begin{figure}
\begin{center}
\includegraphics[width=12cm]{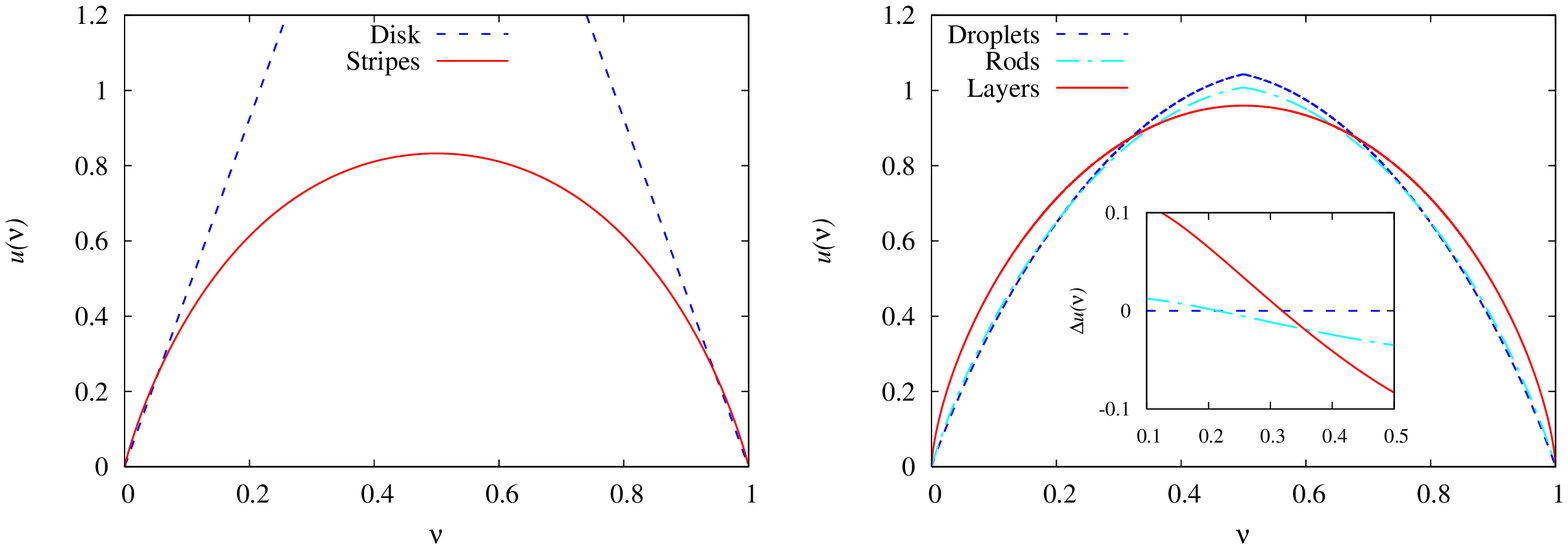}
\end{center}
\caption{The left (right) panel shows the 
function $u(\nu)$ in 
two- (three-) dimensional systems for stripes and disks  (layers,
rods  and droplets).  For the disk, droplet and rod
inhomogeneities, the mixing energy has been computed referring to
$B$-phase inhomogeneities for $\nu<\frac{1}{2}$ while $A$-phase
droplets and rods has been considered for $\nu>\frac{1}{2}$. The inset in the right panel shows the difference in $u$ with respect to the droplet geometry to resolve the crossings between the different morphologies.} 
\label{fig:morphology}
\end{figure}

The low (high) density phase will be termed $A$ ($B$). Defining
 $\tilde f\equiv f/f_0$ the free energy density  can be put as: 
\begin{equation}
\tilde f=\left(1-\nu \right)\tilde f_{A}\left(\phi_{A}\right)+\nu
 \tilde f_{B}\left(\phi_{B}\right)+e_{mix}
\label{eq:fUDA}
\end{equation} 
 where  $\tilde f_{A/B}=(\phi\pm1 )^2$
and $\nu$ indicates the volume fraction of the $B$-phase domains.
 $e_{mix}$ represents the additional energetic cost to form
 inhomogeneities due to the long-range Coulomb interaction and the
 interface boundary energy. For a given geometry it is determined by
 adding the Coulomb  cost and the surface energy cost and optimizing with
 respect to the dimension of the inhomogeneities.   
It can be cast as:
\begin{equation}
e_{mix}=Q^{2/D}_{R} \left(\phi_{B}-\phi_{A}\right)^{2/D} u(\nu)
\label{eq:emixrescale}
\end{equation}
The dependence of the mixed states (MS) free energy upon the
morphology of the domains is stored in the function $u(\nu)$.
 In Fig.~\ref{fig:morphology} we plot $u(\nu)$ for the
different geometries.  These functions have been
evaluated in Refs.\cite{lor01I,ort06} and appear also in the theory of 
diblock copolymers\cite{oht86}.
The  advantage of the UDA approximation is that the mixing energy,
that encloses long-range force effects, can be computed independently
from the specific modeling  of the homogeneous phases free energy.   
These functions are valid whatever form one choses for $f_{A/B}$.

As shown in Fig.~\ref{fig:morphology}  droplet-like or disk inhomogeneities
are preferred for low volume fractions of the minority phase  ($\nu
\sim 0$ or $\nu \sim 1$). On the
contrary stripes and layers  appear for $\nu \sim
\frac{1}{2}$. At intermediate values of $\nu$ in three dimensional 
systems one finds cylindrical rods.  In the diblock copolymer context the control
parameter is the volume fraction $\nu$. Here instead the control
parameter is the global density $\overline\phi$. However 
for $Q^{2}_{R}<<1$, the volume fraction increases nearly 
linearly with the global density and the two parameters are practically
equivalent. 
From Fig.~\ref{fig:morphology} one sees that there will be a series of 
morphological transitions  that
connect droplet states near the homogeneous-MS transitions to the
striped mixed state at $\overline\phi\sim 0$. In the present
approximation they appear as first order however consideration of more
complicated geometries and charge relaxation effects can change this
to a smooth evolution of inhomogeneities that could also
include ``fingering''  and elongation of the domains as in classical
systems\cite{seu95}. The stripe and rod phases are expected to behave as a
glass in quench experiments\cite{sch00} with 
 labyrinth-like patterns.  


Since we are measuring energies in terms of the barrier height, which
represents the characteristic energy gain from phase separation
Eq.~(\ref{eq:emixrescale}) allows to give another
physical interpretation to the frustration parameter:
$$
Q_R^2=\left(\frac{\rm Char.\;mixing\; energy\; cost}{\rm Char.\; phase\; separation\; energy\; gain}\right)^{D}.
$$

In Fig.~\ref{energylambda} we show with dashed lines the typical
behavior of mixed states (MS) free energies  for the $\gamma=1$ model
at different values of $Q^{2}_{R}$. These results are
exact but the results within the UDA approximation at weak frustration
are practically identical\cite{ort07}.  An interesting property of mixed
states is that they present the ``wrong'' curvature; that is, the
electronic compressibility $\partial^{2} f_{e} / \partial \phi^{2}$ is
negative. Generally, this does not imply a thermodynamic instability
since the usual stability condition of positive compressibility must
be formulated for the global neutral system thus including the
background compressibility. Since in frustrated phase separation
analysis, the inverse background compressibility is assumed to be an
infinite positive number (the background density is fixed to the
uniform average value $\overline{\phi}$), it follows from this point of
view that the system is in a stable MS.

An important difference with ordinary phase separation  
resides in the behavior of the
local densities of the domains. In unfrustrated phase separation the
two phases have a constant density independently of the global
density. In frustrated phase separation the local
density of the domains decreases with an increase of the global 
density\cite{lor01I,lor02}. Assuming that the Curie temperature is an
increasing function of the local density, rather than the global
density (controlled by doping), this could explain \cite{lor01II} the
puzzling maximum of the Curie temperature in the three-dimensional
perovskite manganite La$_{1-x}$Ca$_{x}$MnO$_{3}$ at $x=0.35$ Ca doping
\cite{sch95} not predicted by the conventional double-exchange
mechanism \cite{zen51,and55,gen60}.

\section{The strong frustration regime}
\label{sec:stronglambda}

 The results presented above, fully determine the behavior of MS in
the weak frustration regime. By increasing the renormalized Coulomb
frustration in the models
Eqs.~(\ref{eq:eGLrescale}),(\ref{eq:modelrescale}), MS with local
densities close to the reference density  appear. In this
case the behavior of the two models is radically different  leading to
two ``universality'' classes.

We start analyzing the $\gamma=1$ universality class.
As a first approximation we can invoke the UDA keeping in mind that in
this case, since we are far from the Maxwell construction limit, 
the results should be taken with a pinch of salt. Indeed we will see
that special care is needed in $D=2$.  

\begin{figure}
\begin{center}
\includegraphics[width=12cm]{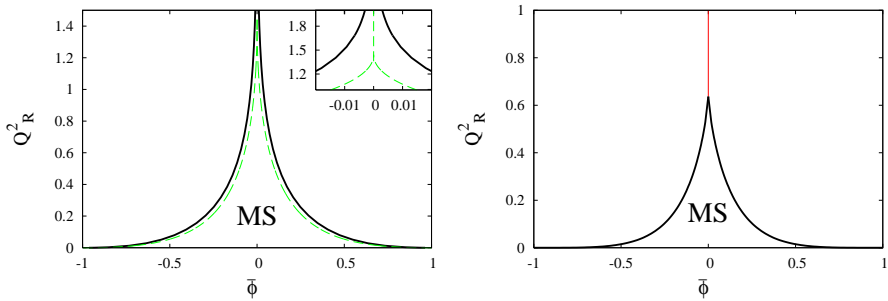}
\end{center}
\caption{The full line in the left (right) panel represents the
exact phase diagram of the $\gamma=1$ model for a smectic striped 
(layered) state in a two- (three-) dimensional system. The dashed line
in the left panel represents the UDA approximation. 
A discrepancy between the two solutions is
  found for $Q^{2}_{R} \sim Q^{2}_{R,c} $ (inset of the left panel) since the UDA
  predicts a critical value of $Q^{2}_{R}$ while the exact model shows
  a logarithmic singularity. In three-dimensional systems the UDA (not shown)
  gives qualitatively the same result as the exact solution.  
  For $Q^{2}_{R}>Q^{2}_{R,c}$  there is a common boundary  
(thin line) between the two homogeneous phases.} 
\label{eq:figphasediagram2d}
\end{figure}

For simplicity we fix the
global density at the reference density $\overline{\phi}=0$  and look for the
energetic stability  
of a striped ($D=2$) or layered ($D=3$) mixed state. By symmetry
$\nu=1/2$ and the  free energy density behaves as
\begin{equation}
\delta \tilde f =  - |\delta \phi|+ \frac14 \delta \phi^{2} +Q^{2/D}_{R} u(1/2) \delta \phi^{2/D}\;\;\;\;(\gamma=1)
\label{eq:deltaf}
\end{equation}   
 where $\delta \phi= \phi_{B}-\phi_{A}$. The first term represents the 
 phase separation energy gain, the second term is an 
energetic cost due to compressibility
 effects and the last term is the UDA mixing energy 
[c.f. Eq.~(\ref{eq:emixrescale})]. The
 condition of stability of mixed states reads $\delta \tilde f<0$. 
In $ D=3$ the last term is dominant at small $\delta \phi$ and
 combined with the linear term produces an energetic barrier
between the homogeneous state ($\delta\phi=0$) and the inhomogneneous
 state   ($\delta \phi\ne 0$). The quadratic term ensures stability
 for large $\delta \phi$. Clearly the transition will be first order
 with a critical frustration given by $Q_{R,c}^2=27/[4 u(1/2)]^3 \sim
 0.47$. Apart form small
 numerical corrections this result coincides with the exact solution
 (c.f. Fig.~\ref{eq:figphasediagram2d}). The different power
 dependence of the phase separation energy gain and the  
mixing energy cost makes the UDA reliable.  

Another important prediction of the UDA is that domains have never all
 linear dimensions much larger than the screening length. This
 important results is quite general and can be derived by simple
 arguments (see below). 
A small difference between the UDA and the exact results arises
 exponentially close to  $(Q^{2}_{R,c},\overline\phi_{c}=0)$ 
where domains with $l_{d}>>l_{s}$ are possible. This region however is
 physically irrelevant since requires an unphysical tuning of the
 density.

For $D=2$ we are in a marginal situation. There is a delicate balance
between the first and the  third terms in Eq.~(\ref{eq:deltaf}) 
which are of the same order.  
The transition looks second order with $Q_{R,c}^2=1/u(1/2)^2\sim 1.45 $. This
result, however, is incorrect. In this marginal case charge relaxation
introduces a correction which unbalances the two terms.  
Fortunately the  model is analytically solvable\cite{jam05,ort07}
 for stripe and layered
structures which are the expected morphologies close to
$\overline\phi=0$.  In
Fig.~\ref{eq:figphasediagram2d} we show the exact phase diagrams for
these particular morphologies in two- and
three-dimensional systems. In $D=2$ the transition lines diverge
logarithmically at $\overline\phi=0$ and there is no upper limit for the mixed
state.  Therefore a 
direct first-order phase transition between homogeneous phases is not 
possible and there is always an intermediate phase.

\begin{figure}
\begin{center}
\includegraphics[width=7cm]{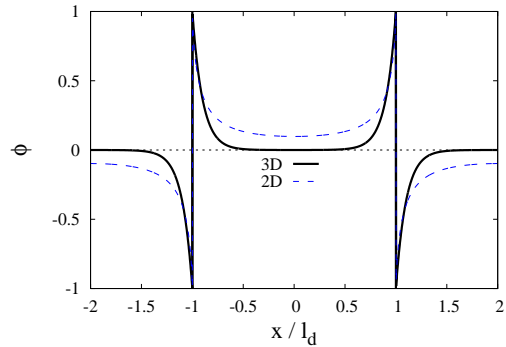}
\end{center}
\caption{Behavior of the charge density modulation at the reference
  density $\overline\phi=0$ for a cut perpendicular to stripes in
  two-dimensional systems and layers in three-dimensional systems and
  domain width $W=20 l_{s}$.} 
\label{fig:chargedensity}
\end{figure}

 The same result has been first obtained in this region of the phase diagram by
neglecting the compressibility term in the free energy $\propto \phi^{2}$
\cite{jam05}. Indeed,  in the strong frustrating regime the typical
size of 2D domains is exponentially larger than the screening length
and thus the physics is determined by the slow power-law relaxation of
the charge density  (Fig.\ref{fig:chargedensity}). The screening
length plays the role of a short-length cutoff that removes the
unphysical divergence of the charge density at the domain boundaries
arising at $\beta=0$. 

For a practical point of view strongly frustrated 2D mixed state
can be quite difficult to observe.  It appears in the
exponentially narrow range of densities 
$$|\overline{\phi}|< e^{- Q^{2}_{R}/4}$$ and therefore it
requires an enormous accurate control of the density. 
It may be possible that this is
achieved to some extent in ruthanates, where the control parameter is the
magnetic field which allows for considerable fine tuning and the
system is ultraclean. 

Another difficulty is that the inverse electronic compressibility  
$$\kappa_{MS}^{-1}\propto \frac{- \kappa^{-1} e^{ Q^{2}_{R}/4}}{\sqrt{1-\left(e^{Q^{2}_{R}/4} \overline{\phi}\right)^{2}}}$$ 
is exponentially large and negative in all the mixed state stability range 
and negatively diverges with $1/2$ critical exponent at the transition. 
As stated in Sec.~\ref{sec:weaklambda}, in frustrated phase separation models the background
is assumed incompressible ($\kappa_{b}\equiv 0$) but real systems will have a
small finite background compressibility $\kappa_{b}>0$. This will
lead to a volume instability analogous to the volume instability of
Cerium metal\cite{lor01I,bus03}.  

In some cases, when there is a large
separation of energy scales, the background may behave as nearly
incompressible. This can occur in  ruthanates 
where the relevant electronic phenomena occur  at temperatures
below a tenth of a kelvin to be compared with the melting temperature
of the material of the order of hundreds of kelvin. In the case of a  
2D electron gas with a ionic background there is also a similar separation of
energy scales. Even more, from an elastic point
of view, the background is three dimensional and therefore behaves as
practically incompressible. In field-effect-transistors where the
background is made of mobile charges, the situation is quite different   
and the background compressibility can not be neglected\cite{spi03}.

Differences arising between two- and three-dimensional systems are
mainly due to the different behavior of the charge density profile
inside the domains. Phase
separation energy gain stems from the region where the electronic
density is significantly different from its average value\cite{lor01I}. 
In two-dimensional systems the power-law behavior of the charge
density allows to gain phase separation energy even far from the
boundaries.  On the contrary,
in three-dimensional systems  the charge density decays exponentially
from the domain boundary on the scale of the Thomas-Fermi screening
length. Thus the system gains phase separation energy in a range
of the order of the screening length from the domain
boundaries. Regions far from the boundaries produce 
an exponentially small energy gain. Thus if phase separation is
favorable the system adjust itself in such a way to eliminate all
these regions. Domains satisfy a maximum size rule such that every
point of the domain is at a distance of the order of the screening
length or smaller from the boundaries. This rule proves to be quite
general and is independent of the model.   
The maximum size rule allows for arbitrary large
inhomogeneities in two-dimensional systems since one of the dimensions
is already smaller than $l_{s}^{3D}$.


Now we discuss the case of $\gamma=2$ at strong frustration.
The separation of length scales discussed in Sec.~\ref{sec:weaklambda}
is no longer valid, the UDA becomes unreliable and the effect of
smooth interfaces plays a prominent role. 

Computing the static response to an external field in momentum space
one obtains the following charge susceptibility: 
$$ \chi (q)= \left[q^{2}-2+6 \overline{\phi}^{2}+\frac{Q^{2}_{R}}{2}
  v(q)\right]^{-1}. $$ 
We remind that  the system is considered to be $D$-dimensional but
  embedded in the usual three dimensional Coulomb
  interaction. The Fourier transform reads: 
$$v(q)=\frac{2^{D-1} \pi}{|q|^{D-1}}.$$
$\chi$ diverges for $D=2,3$ at a characteristic finite
wavevector $q_{c}$ on an instability line
$Q^{2}_{R}(\overline{\phi})$.
Above $Q^{2}_{R,c}=1/(2 \pi)$ in 3D and $Q^{2}_{R,c}=(4 \sqrt{2})/(3 \sqrt{3} \pi)$ in 2D,  the system is always homogeneous. On entering in the unstable region a
sinusoidal charge density wave (SCDW) occurs. Thus
inhomogeneities are quite different form the mesodomains  of the weak coupling
regime (Sec.~\ref{sec:weaklambda}). The crossover between this two
regimes will be discussed elsewhere \cite{ort07bis}.

A related mechanism has been proposed in
cuprates predicting charge ordering instabilities
and other anomalous properties in accord with experiment\cite{cas95b}. 
In classical systems this  is ofter referred as the ``microphase''
separation transition \cite{seu95,mur02}.

\section{Conclusions}
\label{conc}
In this work we reviewed the main aspects of frustrated phase
separation in charged systems considering two kind of compressibility anomalies that
generally occur in strongly correlated systems. For the different
coarse-grained models that have been introduced in the literature, 
the outcome of long-range forces can
be measured by a dimensionless parameter that defines the amount of
frustrating effects. Frustration tends to reduce the range of
density where a mixed state appears. Thus uniform phases are possible
at densities where in the absence of long-range forces, phase separation would
occur.  This situation is in accord with  thermodynamic 
measurements \cite{eis92,eis94} of the uniform 2DEG where one finds 
stability of the uniform phase with negative compressibility. 

In the presence of frustration the mixed state consists of domains
of mesoscopic size with various geometries depending on the control
parameters. A sequence of morphological  transitions occurs that has a 
strong similarity to those in other systems \cite{seu95,oht86}.
  
When frustrating effects are a small perturbation, a unified
description of MS is allowed and a simple approximation in which the
density inside the domains is assumed constant, gives very accurate
results. 

At strong frustration, the properties of the mixed state strongly
depends upon the 
particular anomaly in the compressibility  
and one can define two different universality classes.   
For systems with a cusp singularity in the electronic
compressibility ($\gamma=1$) the system dimensionality
plays a key role in the charge segregation mechanism. A critical
frustration $Q_{R,c}$, above which mixed states are not possible, exist
only in $D=3$.   

In systems with a negative electronic compressibility
region ($\gamma=2$) a critical value of the frustration exist
$Q_{R,c}$ for both  $D=2,3$. Close to $Q_{R,c}$ inhomogeneities 
are SCDW.

The maximum size rule  says that 
domains  can not have all linear dimensions  much larger than the
screening length independently of $Q_R$. 
Thus a necessary condition for the applicability of
a coarse grained treatment is that the three dimensional 
screening length must be much greater than the typical microscopic
lengths like the interparticle  distance and the lattice constant.  
Therefore mesoscopic domains can
be expected in systems with small compressibility as bad metals or
systems close to metal-insulator transitions or 
systems with very anisotropic electronic properties (i.e. are nearly 
insulating in one direction). Interestingly most of
the materials where mesoscopic domains have been found have these
characteristics.

\end{document}